\documentclass[sigconf]{acmart}

\AtBeginDocument{%
  }

\acmPrice{15.00}
\acmISBN{978-1-4503-XXXX-X/18/06}
\copyrightyear{2023}
\acmYear{2023}
\setcopyright{rightsretained}
\acmConference[MICRO '23]{56th Annual IEEE/ACM International Symposium on Microarchitecture}{October 28-November 1, 2023}{Toronto, ON, Canada}
\acmBooktitle{56th Annual IEEE/ACM International Symposium on Microarchitecture (MICRO '23), October 28-November 1, 2023, Toronto, ON, Canada}
\acmDOI{10.1145/3613424.3614308}
\acmISBN{979-8-4007-0329-4/23/10}

\usepackage{graphicx}
\usepackage[caption=false]{subfig}

\begin{document}

\title[Efficiently Enabling Block Semantics and Data Updates in DNA Storage]{Efficiently Enabling Block Semantics and Data Updates in DNA Storage}

\author{Puru Sharma}
\email{puru@u.nus.edu}
\orcid{0009-0000-5854-1086}
\affiliation{%
  \institution{National University of Singapore}
  \country{}
}
\author{Cheng-Kai Lim}
\email{limchengkai@u.nus.edu}
\orcid{0000-0003-4640-9883}
\affiliation{%
  \institution{National University of Singapore}
  \country{}
}
\author{Dehui Lin}
\email{e0203126@u.nus.edu}
\orcid{0009-0005-3437-2683}
\affiliation{%
  \institution{National University of Singapore}
  \country{}
}
\author{Yash Pote}
\email{yashp@comp.nus.edu.sg}
\orcid{0000-0002-5060-684X}
\affiliation{%
  \institution{National University of Singapore}
  \country{}
}
\author{Djordje Jevdjic}
\email{djolent@gmail.com}
\orcid{0000-0002-3341-9364}
\affiliation{%
  \institution{National University of Singapore}
  \country{}
}

\renewcommand{\shortauthors}{Sharma, et al.}

\begin{abstract}
We propose a novel and flexible DNA-storage architecture, which divides the storage space into fixed-size units (blocks) that can be independently and efficiently accessed at random for both read and write operations, and further allows efficient sequential access to consecutive data blocks. In contrast to prior work, in our architecture a pair of random-access PCR primers of length 20 does not define a single object, but an independent storage partition, which is internally blocked and managed independently of other partitions. We expose the flexibility and constraints with which the internal address space of each partition can be managed, and incorporate them into our design to provide rich and functional storage semantics, such as block-storage organization, efficient implementation of data updates, and sequential access. To leverage the full power of the prefix-based nature of PCR addressing, we define a methodology for transforming the internal addressing scheme of a partition into an equivalent that is PCR-compatible. This allows us to run PCR with primers that can be variably elongated to include a desired part of the internal address, and thus narrow down the scope of the reaction to retrieve a specific block or a range of blocks within the partition with sufficiently high accuracy. Our wetlab evaluation demonstrates the practicality of the proposed ideas and a 140x reduction in sequencing cost and latency for retrieval of individual blocks within the partition.
\end{abstract}

\maketitle

\section{Introduction}
\label{sec:intro}

The rapid improvements in performance and cost of DNA synthesis and sequencing methods have led to increased interest in the use of DNA as a durable and compact medium for data storage, with a large spectrum of available chemical tools that enable efficient data access and manipulation of in-DNA data. A number of DNA storage architectures have been proposed~\cite{church:next, goldman:towards, grass:robust, yazdi:rewritable, yazdi:portable, organick:random, bornholt:dna, banal:random, lin:managing}, with numerous encoding schemes that seek to improve the resilience against various types of complex errors, maximize the information density, and allow for efficient data access. Computational primitives such as \textit{similarity search} have been designed on top of DNA storage~\cite{bee:molecular, stewart:content}, micro-fluidic chips and runtimes have been developed to automate wetlab protocols in DNA storage~\cite{willsey:puddle}, and simple but fully automated end-to-end DNA storage systems have already been demonstrated~\cite{takahashi:demonstration}.

One of the most notable features of DNA storage is that it can provide random access at nearly constant latency~\cite{yazdi:rewritable, organick:random, bornholt:dna}, regardless of the amount of data in the storage system. This is often achieved through the use of the ~\emph{Polymerase Chain Reaction} or PCR, which is one of the most important and well-understood chemical reactions with a wide range of routine uses in biochemistry. PCR is a parameterizable reaction, the parameters being two short DNA sequences, called \emph{primers}. For optimal PCR performance, the length of each primer is typically fixed at 20 characters~\cite{yazdi:rewritable, organick:random, bornholt:dna, tomek:promiscuous}. When two primers are inserted into a sample containing a DNA storage system, running a single cycle of PCR will duplicate all those DNA fragments in the sample that begin with the first primer (\emph{forward primer}) and end with the second primer (\emph{reverse primer}). After several PCR cycles, the targeted sequences will multiply exponentially and become predominant in the sample. At that point, the target DNA sequences can be easily read or \textit{sequenced} at low price, without reading all of the sequences from the original sample.~\footnote{PCR is also applicable to RNA, for example in PCR COVID-19 tests, where the appropriate primers would correspond to stable and well-known parts of the viral RNA.} 

Most of the state-of-the-art DNA storage architectures are designed around the mentioned PCR mechanism that provides random access. Logically, they are organized as key-value object stores~\cite{bornholt:dna, organick:random, lin:managing}, in which a pair of primers define the key, and the \emph{arbitrarily sized} value is stored in molecules which are tagged with the same pair of primers. The layout of a DNA molecule in this architecture is shown in Figure~\ref{fig:back}a. Note that a large object can be stored across many molecules containing the same pair of primers, all of which are retrieved during PCR with sufficient uniformity~\cite{organick:random}; a part of each molecule is thus reserved to uniquely identify the molecules and re-establish the order between them in software, and we refer to it as the \textit{index} or \textit{internal address}. An object that spans \textit{N} molecules requires ${\log_4 N}$ bases for indexing. 

While there are $4^{40} = 2^{80}$  unique combinations of letters that could be used to define a pair of primers, most of them are not appropriate for PCR. First, PCR primers must have \textit{balanced GC content}, which means that the aggregate percentage of G and C characters in the whole primer sequence should be close to 50\%~\cite{yazdi:rewritable, organick:random}. Second, all primers used in the same DNA sample must be significantly different from each other in Hamming distance to avoid amplification of unwanted data~\cite{yazdi:rewritable, organick:random}. This constraint on minimum pairwise distance between each pair of primers turns out to be a major problem~\cite{tomek:promiscuous, winston:combinatorial, tomek:driving}, as the largest set of primers found so far to meet such requirements contains only between $\sim$1000-3000 primers, depending on the distance thresholds assumed~\cite{yazdi:rewritable, organick:random}. This allows us to chemically distinguish between only about 1000 different objects in a DNA pool through PCR~\cite{organick:random}. To illustrate the implications of this limitation, let's assume that we have a DNA pool containing 1TB of data, which is comparable to an average storage disk. Given that we can chemically tag the data with at most 1000 mutually compatible primer pairs, the actual unit of random access is going to be around 1GB on average. As a result, if a user wants to retrieve 1MB of data, they will have to first amplify (through PCR) and then sequence a whole 1GB of data, with $\sim$99.9\% of the sequencing output representing irrelevant data. Consequently, $\sim$99.9\% of the sequencing latency and cost is being wasted. The user would then use software tools to extract the desired 1MB of data out of the sequenced 1GB, discarding most of the sequenced data~\cite{organick:random, bornholt:dna}.  Unfortunately, simply using longer primers would not alleviate the problem. According to the available analytical models, the number of compatible primers scales approximately linearly with the primer length~\cite{organick:random}. We empirically tested the same methodology~\cite{organick:random} on primers of length 30, and we managed to find only around 10K primers that meet the requirements, which hardly justifies the additional synthesis cost of the longer primers, paid for every individual molecule.

While it is quite impressive that, unlike any other device, DNA can natively support the object storage semantics, we find that such semantics is the root cause of the above problems. By allowing the objects to be arbitrary in size, reliable retrieval of a small object in a sea of giant ones becomes challenging and requires high Hamming distance between the keys, as well as many cycles of PCR. By designing the keys (i.e., primers) of all objects to be as distant from each other as possible, the number of addressable objects is drastically reduced, which in turn significantly limits the possibility to build a functional storage system. Furthermore, the object-based design results in a flat key-value store architecture, without any logical order or distance metric between the objects, limiting sequential access to a single object and essentially degenerating it to a random access. Apart from the above problems, the object storage semantics is largely unnecessary, as any storage system, including object storage, can be implemented on top of a block device.



This work proposes a novel and flexible DNA storage architecture that offers the block storage semantics, where each block can be independently read and written to, and a group of consecutive blocks can be efficiently retrieved. In contrast to prior work, in our architecture, a pair of primers of length 20 do not define a single object, but an independent storage \textit{partition}, which is internally blocked and managed independently with its own index structure. We make the observation that, while the number of mutually compatible primer pairs is limited, the internal address space available to any pair of primers (i.e., partition) is virtually unlimited. We expose and leverage the flexibility with which this address space can be managed to provide rich and functional storage semantics. Furthermore, to leverage the full power of the prefix-based nature of PCR addressing, we define a methodology for transforming an arbitrary indexing scheme into a PCR-compatible equivalent. This allows us to run PCR with primers that can be variably extended to include a desired part of the index, and narrow down the scope of the reaction to retrieve a specific block or a range of blocks with sufficiently high accuracy.

In this paper we make the following contributions:
\begin{itemize}
\itemsep0em 
    \item We expose the flexibility and constraints in the management of the internal address space for any pair of primers, and show that significant functionality can be achieved at negligible losses in information density. We show that this flexibility can be leveraged to implement the block storage semantics. 
    \item We define a methodology for designing PCR-compatible indexes to enable random access with primers that can be elongated to include a variable portion of the index, called elongated primers, enabling sufficiently reliable random access to individual blocks and sequential access to consecutive blocks.
    \item We provide efficient support for data updates in DNA storage by organizing the partitions similarly to version control systems; this style of updates obviates the need for complex chemical edits of DNA molecules and avoids their limitations, and allows for efficient retrieval of updated data, while adhering to the block storage semantics. 
    \item Our wetlab evaluation using the state-of-the-art DNA storage architecture~\cite{organick:random} demonstrates the practicality and precision of our random access with elongated primers, in which we were able to reduce the sequencing costs by $\sim$140x. We further demonstrate the practicality of the proposed data update approach, by synthesizing several DNA update \textit{patches} with carefully engineered addresses, and carefully mixing them with the original data. We then show that \emph{in one round-trip to DNA storage} we can precisely retrieve (and later successfully decode) a small part of data that experienced updates, with minimal sequencing of unrelated data and unrelated updates. 
\end{itemize}

The rest of the paper is organized as follows. Section~\ref{sec:back} covers the background information about DNA storage. Section~\ref{sec:internal} analyzes trade-offs in internal address space management of a single partition.
Section~\ref{sec:elongated} describes the random access to individual blocks and sequential access to consecutive blocks. Section~\ref{sec:update} describes existing approaches to enabling updates in DNA storage, discusses the trade-offs involved, and presents our solution that aligns with the proposed block semantics. Sections~\ref{sec:methodology} and ~\ref{sec:evaluation} describe our experimental methodology and the results of our wetlab experiments, respectively. Section~\ref{sec:decod} describes our data decoding algorithm. We discuss the related work in Section~\ref{sec:disc} and conclude in Section~\ref{sec:conclusion}.

\section{Background}
\label{sec:back}

\subsection{DNA Storage Basics}
\begin{figure}
	\includegraphics[width=1\columnwidth]{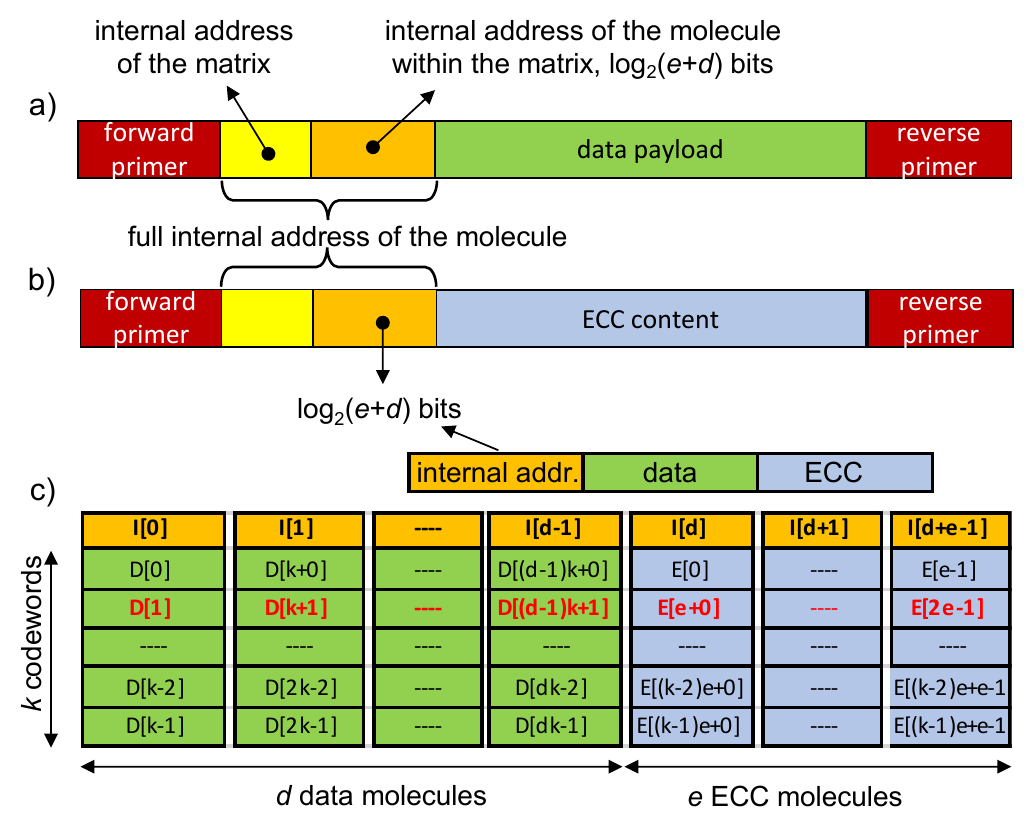}
	\caption{a) Structure of a DNA molecule in DNA-based storage~\cite{organick:random}, b) layout of an encoding unit with an outer ECC code~\cite{organick:random}, and c) layout of data and ECC molecules within an encoding unit, layed out as a matrix.}
	\label{fig:back}
\end{figure}

The key enabler of DNA-based storage systems is a chemical process called \emph{artificial DNA synthesis}, widely available as a commercial service. The process can create an arbitrary sequence of \{A, C, G, T\} bases, which may or may not have biological meaning. While natural DNA molecules in living cells tend to be huge and carry all of the organism's genetic information, the artificial molecules produced by the current synthesis technologies are limited to a few hundred bases in length. Although it is possible nowadays to synthesize DNA molecules that are over 1000 bases long~\cite{yazdi:rewritable, yazdi:portable}, the quality assurance costs increase sharply beyond a few hundred bases, and the yield decreases significantly. Given that a DNA molecule that is 300-base long can store at most 75 bytes of information, storing large files requires breaking them down into smaller pieces that fit into shorter molecules. Some form of an \emph{internal address}, also called an \emph{index}, must also be added to each piece of data, i.e., it must be embedded into every molecule, to allow for eventual reassembly of the original data from the pieces~\cite{heckel:fundamental, organick:random, bornholt:dna}. Figure~\ref{fig:back}a shows the structure of a DNA molecule in the state-of-the-art architecture~\cite{organick:random} with an internal address.

\subsubsection{Encoding}
To be stored in a DNA format, pieces of binary data must be first encoded into a series of \{A, T, C, G\} bases (nucleotides) or  using one of many available coding schemes. Some coding schemes tend to sacrifice the coding efficiency to adhere to rules that help with the success of the chemical processes in the pipeline. For example, they may try to prevent the occurrence of long stretches of \emph{homopolymers} (repeating bases, e.g., AAAA) in order to make the sequencing process easier~\cite{organick:random}. Other coding schemes try to balance the GC-content in order to make the DNA synthesis more successful~\cite{yazdi:rewritable, yazdi:portable}. This type of coding is known as \emph{constrained} coding, and is employed by most of the early work on DNA storage~\cite{bornholt:dna, c:dube, c:banerjee, c:erlich, c:press, c:nguyen, c:park}. In contrast, \emph{unconstrained} coding does not seek to exclude any particular sequences, and it often employs simple data randomization, which ensures that long homopolymers occur with low probability and the GC-content is balanced on average~\cite{embracing}. Unconstrained coding significantly increases the coding density, while relying on conventional error correcting codes to handle all error types. In this work we employ unconstrained coding, assuming a simple mapping of two bits per base, which achieves the maximum information density, while we handle all error types efficiently using outer Reed-Solomon ECC codes~\cite{organick:random, lin:managing}. This approach has been shown to lead to much higher information density for practical ranges of error rates~\cite{embracing}. However, to encode the internal addresses within every molecule (more specifically, the yellow part in Figure~\ref{fig:back}a), we use a more complex constrained coding scheme that allows us to implement the block storage semantics (Section~\ref{sec:elongated}).

Once a file has been split into pieces and encoded into DNA strings, a pair of special identifying sequences called \emph{primers} are added to the beginning and end of each string. These primers constitute a chemical tag that logically groups related molecules together, as shown in Figure~\ref{fig:back}a, allowing for random access. The tagged sequences are sent for a commercial synthesis service, whereby millions of physical copies of each DNA string are synthesized together and stored in the same DNA pool (e.g., in a test-tube). 

\subsubsection{Data Retrieval}
To access a file stored in a DNA pool, the molecules containing individual pieces of the desired file need to be read through a process called \textit{DNA sequencing}. To isolate the target molecules to be sequenced, the PCR reaction is used to selectively \textit{amplify} (i.e., exponentially multiply) the molecules containing the file pieces. These molecules are isolated using their primers. Once the target molecules are amplified, they are sequenced using one of many available sequencing technologies. The sequencing produces many DNA strings, often called  \textit{reads}. The synthesis, storage, wetlab manipulation, and sequencing processes often result in errors that manifest themselves in the final reads when compared to the originally encoded DNA molecule~\cite{simulating}. The average number of reads per original synthesized molecule is called \emph{sequencing coverage} or \emph{sequencing depth}. The higher the coverage for a molecule, the easier it is to reconstruct it from these reads, but also the higher the cost of sequencing.


The obtained reads that contain the correct primers are then clustered based on similarity such that each cluster ideally consists of all reads originating from the same encoded DNA molecule. The similarity metric typically used for clustering is the Levenshtein (edit) distance~\cite{rashtchian:clustering, lin:managing, organick:random}, defined as the minimum number of insertion, deletion or substitution operations required to convert one string to another. Each cluster contains noisy copies of the same original DNA string, which is then extracted from each cluster using one of many consensus finding algorithms~\cite{organick:random, sabary:reconstruction, bhardwaj:trace, yazdi:portable, lin:managing, sergey:trellis}.

\subsubsection{Decoding and Error Correction}
After clustering and finding the most probable original strand for each cluster, the obtained strands are decoded back into binary data. Using the internal address information stored in each strand, the binary data is used to recreate the original file by reordering the pieces. Any errors left over from other steps in the pipeline are corrected using error-correction codes, typically outer Reed-Solomon or LDPC codes~\cite{organick:random, grass:robust}. These schemes group a larger number of molecules (tens of thousands~\cite{organick:random}) into encoding units, to enable erasure-coding in case of losses of entire molecules, and to amortize the cost of error correction over a larger set of data. For example, the state-of-the-art architecture~\cite{organick:random} treats all DNA molecules as columns in a matrix, as shown in Figure~\ref{fig:back}c. Separate DNA molecules are then created as an external ECC code, such that one codeword represent a row in this matrix (the row in red in Figure~\ref{fig:back}c). While this type of coding achieves high information density, it creates strong inter-molecular dependencies across thousands of molecules, which make data edits extra challenging. 

\begin{figure}
	\includegraphics[width=1\columnwidth]{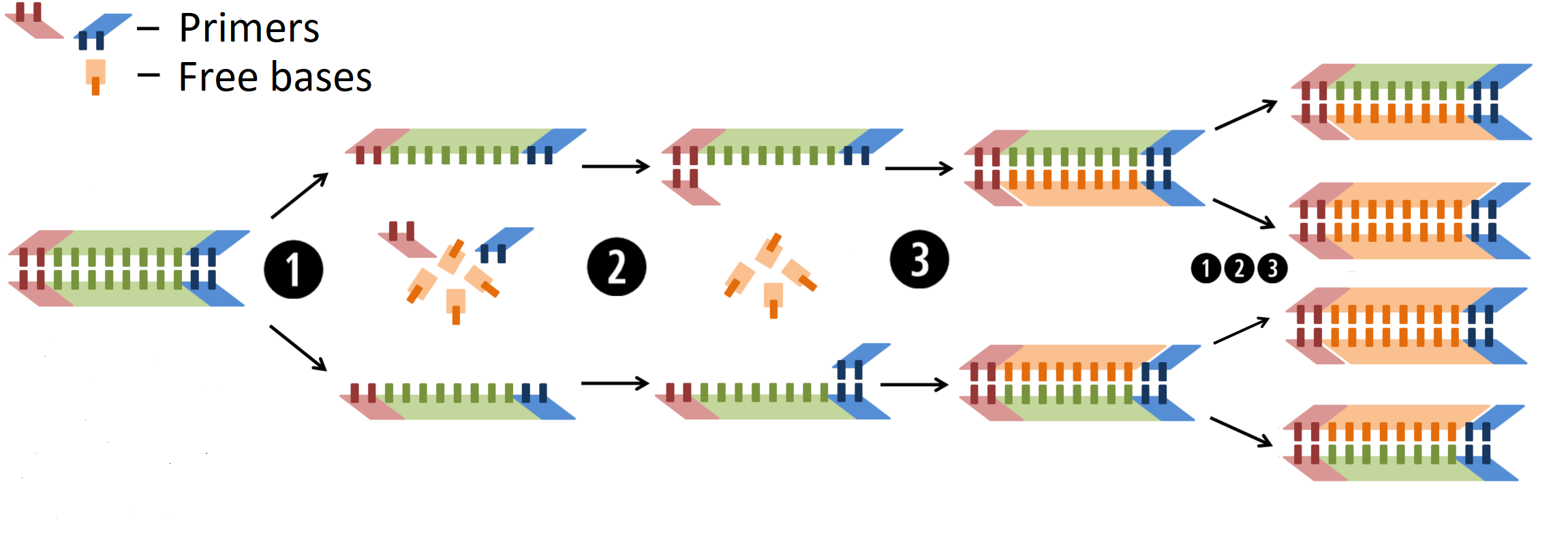}
	\caption{Three steps of a PCR cycle that double the number of desired molecules.}
	\label{fig:pcr_steps}
\end{figure}

\subsubsection{PCR}
PCR is a cyclical reaction that doubles the number of the desired molecules (also known as \emph{target} molecules) in every cycle, as illustrated in Figure~\ref{fig:pcr_steps}. All reagents are put into a test-tube, which is then closed and exposed to different temperatures in a programmable manner. A PCR cycle contains three phases, defined by the temperature: 1) denaturation, which happens at 95$^{\circ}$C and breaks a double-stranded molecule into two single-stranded ones, 2) annealing, which is the binding of primers to the target molecules, and happens at the temperature of 48-68$^{\circ}$C, 3) extension, which creates a double-stranded molecule between two primers and happens at 68-72$^{\circ}$C. 

The optimal primer length for PCR is considered to be around 20, which provides the optimal conditions for binding (\emph{annealing}) of the primers to the target molecules~\cite{dieffenbach:general}; primers of length 20 anneal at 50$^{\circ}$C and do so efficiently, whereas longer primers anneal less efficiently and at higher temperatures. However, PCR with primers of length 40+ is routinely done in biological applications~\cite{innis:pcr, dieffenbach:general} and in most sequencing protocols~\cite{basu:pcr}, and the maximum length of the primers was recently found to be above 100~\cite{tian:length}. Therefore, in a DNA storage system, one could in theory retrieve individual molecules using PCR, as each molecule contains a unique prefix consisting of its primer and the full internal address; the full prefix would be used as the forward PCR primer. Unfortunately, the internal addresses can be very similar to each other, they don't have a balanced GC content, and they may have a very long stretches of repeated bases (homopolymers), and PCR with such primers easily fails~\cite{dieffenbach:general, basu:pcr}.

\subsection{Existing Data Update Mechanisms}
DNA-based data storage is largely assumed to be read-only~\cite{bornholt:dna} due to impracticality of updating already created DNA molecules. However, multiple techniques for in-situ alterations of DNA molecules have been proposed as a mechanism for rewriting data stored in DNA and have been tested in simple proof-of-concept setups~\cite{yazdi:rewritable, lin:dynamic, pan2022rewritable}.

Direct edits of DNA molecules were proposed as the first method to support updates in DNA, and the proof of concept was successfully demonstrated on a single long DNA molecule~\cite{yazdi:rewritable}. This method effectively cuts out a fragment of DNA that needs to be updated and \textit{pastes} a new one using overlap-extension PCR. More recently, strand displacement technology has been used as an edit method~\cite{lin:dynamic}, with multiple simultaneous single-molecule edits in the same cycle. Both processes are chemically complex and could be risky, as they may cause accidental modifications of other data in sufficiently large DNA pools. 

However, a bigger problem with both technologies is that they are limited to updates within a single molecule and can be applied only when the size of the data does not change as a consequence of the update. The state-of-the art organizations of DNA storage~\cite{organick:random} assume that data is split across many molecules, interleaved in multiple ways across an even bigger set of DNA molecules (e.g., to efficiently implement error-correcting and erasure codes, like in Figure~\ref{fig:back}c). This creates significant inter-molecular data dependencies that make the use of molecular editing impractical without breaking the structure of surrounding data, or even impossible when the size of the data changes. 

Instead of encoding data as nucleotides in DNA strands, recent work has proposed storing data as nicks in the DNA backbone~\cite{tabatabaei2020dna,pan2022rewritable}. One advantage of this method is that these nicks can be repaired, and then reapplied, making the data storage system rewritable, although one must erase the data in the entire pool in order to rewrite it. However, adding nicks to DNA strands means that one cannot perform PCR on them, sacrificing random access and the ability to read data using next-generation sequencing. Importantly, the storage density of this approach is 50-fold lower compared to storage of data as nucleotides~\cite{tabatabaei2020dna}. Given the above, such an approach is more appropriate for storing small metadata associated with the entire DNA pool as a whole, rather than the data itself.



\section{Managing Internal Address Space}
\label{sec:internal}

\begin{figure}
	\includegraphics[width=1\columnwidth]{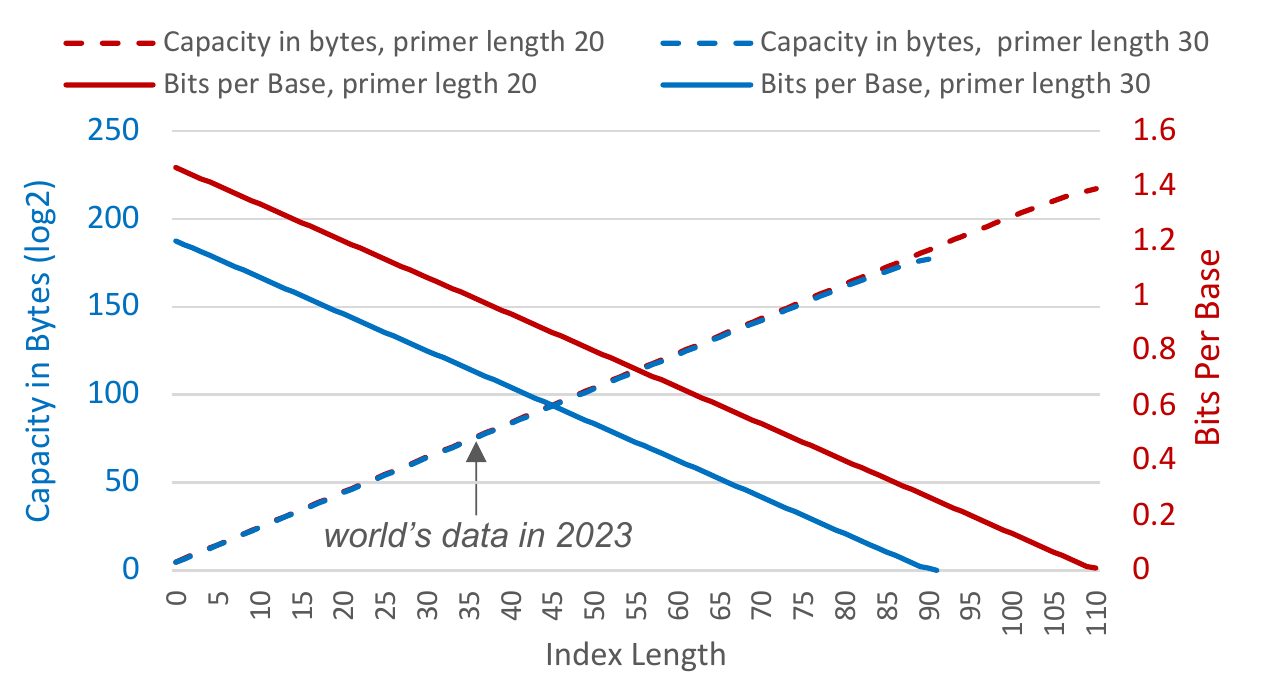}
	\caption{Storage capacity (bytes) of a single partition (blue) and information density (red) as a function of the index length. Dashed lines correspond to design with primers with 30 bases.}
	\label{fig:internal}
\end{figure}

To understand the trade-offs involved in the management of the internal address space, let us quantify the storage capacity and information density of a DNA storage partition defined by a pair of primers of length 20. Let's assume that the length of the DNA strands is 150, as in our wetlab experiments. Figure~\ref{fig:internal} shows the storage capacity (blue, left y-axis) and the information density (red, right y-axis) as a function of the index length \textit{L}. Note that the maximum storage capacity of $2^{217}$B is achieved when the entire available portion of the strand is used for indexing; in that case there is no space for data within the molecule, but the presence of a molecule is treated as 1, and the absence as 0, and this single-bit value is assigned to the internal address, of which there are $4^{110}$=$2^{220}$. This design however has extremely low information density: only one bit per strand of 150 bases.

In contrast, the density is the highest when there is only one molecule which requires no index at all, so the entire strand can be used for data. However, this density is achievable only for tiny objects. Also note that designs with primers of length 30 (dashed lines), although achieving lower storage capacity, still have enormous capacity that greatly surpasses the current amount of data in the world. While longer primers do reduce the information density significantly, this loss diminishes linearly with longer strand length.

\subsection{Partition Architecture}
The address space AAA...AAA to TTT...TTT with an index of length $L$ can be represented as a prefix tree, shown in Figure~\ref{fig:tree}a, which represents the hierarchy of addresses within a partition. Every non-leaf node in this tree has four edges labelled A, C, G, T, in that order. Every \emph{leaf} represents one DNA strand, whose full internal address is determined by the path from the root to the leaf. For example, the left-most leaf would be AAA....AAA ($L$ characters), and the right-most leaf would be TTT...TTT. 

When this tree is balanced and full, i.e., when all DNA strands have the same index length and all indexes are present, then the maximum information density is achieved~\cite{heckel:fundamental}. In such a design, an index of length $L$ covers addresses from AAA...AAA to TTT...TTT, which are treated as a 1D array of $4^L$ fixed-capacity storage units~\cite{bornholt:dna, organick:random, lin:managing}. The size of the unit corresponds to the payload of one molecule. An important observation about this architecture is that \emph{any contiguous byte-range can be statically mapped to a contiguous index-range and vice versa}, just like in block storage. A contiguous index-range, in return, can be precisely described with a few prefixes, or less precisely with their longest common prefix. For example, range AAA to AGT can be precisely described with the following set of prefixes: AA, AC, AG. The longest common prefix is A. However, the set of data covered by prefix A also includes AT, on top of the desired range AAA to AGT.

An important implication of this observation is that any contiguous range of bytes within a partition, which could correspond to an object of any size or a set of contiguously stored objects, could be retrieved quite precisely with a single PCR, if the primers were extended to include a part of the index, assuming that the index complies with PCR primer design constraints. In Section~\ref{sec:elongated} we discuss how to make indexes PCR-compatible through sparse encoding of indexes. 
Also note that in order to improve the efficiency of sequential accesses, a set of files could be mapped onto the partition in a manner that tries to optimally align the files to nodes in the prefix tree, which we leave for future work.

\subsection{Concentration constraints}
To manage the cost of sequencing, every strand in DNA storage should ideally be represented in approximately equal concentrations. Otherwise, highly concentrated strands will be sequenced at much higher coverage than needed, while other strands will need much more sequencing to be represented in the readout, wasting the sequencing resources. Furthermore, for our random block access to work, it is essential that the desired sequences are amplified stronger than other sequences that may be wrongly amplified due to the similarity in the index. Otherwise, strands with an index similar to the desired index may become dominant. Even worse, PCR may overwrite their index to the desired index if the indexes are too similar, resulting in its exponential amplification, a situation called~\emph{mispriming}. In that case, we may have multiple data candidates that present themselves with the desired index, and we may not be able to decide correctly which one is the actual target. For example, assume two strands with very similar indexes I1 and I2, and payloads P1 and P2, and we want to amplify strand (I1, P1). The dominant strands in the outcome could be (I1, P1), which is correct, and (I1, P2), which is a false positive. To ensure the dominance of the desired strands in the presence of similar targets, it is sufficient to ensure that the closest targets in the pool are not present in higher concentration compared to the actual target~\cite{basu:pcr, mispriming}. Applying this rule transitively implies that all the nodes in the same level of the index tree map to the similar number of DNA strands and in similar concentrations. This is trivially ensured in the our partition architecture, however, we need to ensure that data updates do not compromise this balance of strands and their concentrations.
\section{Random Block Access}
\label{sec:elongated}

\begin{figure}
\begin{center}

	\includegraphics[width=\columnwidth]{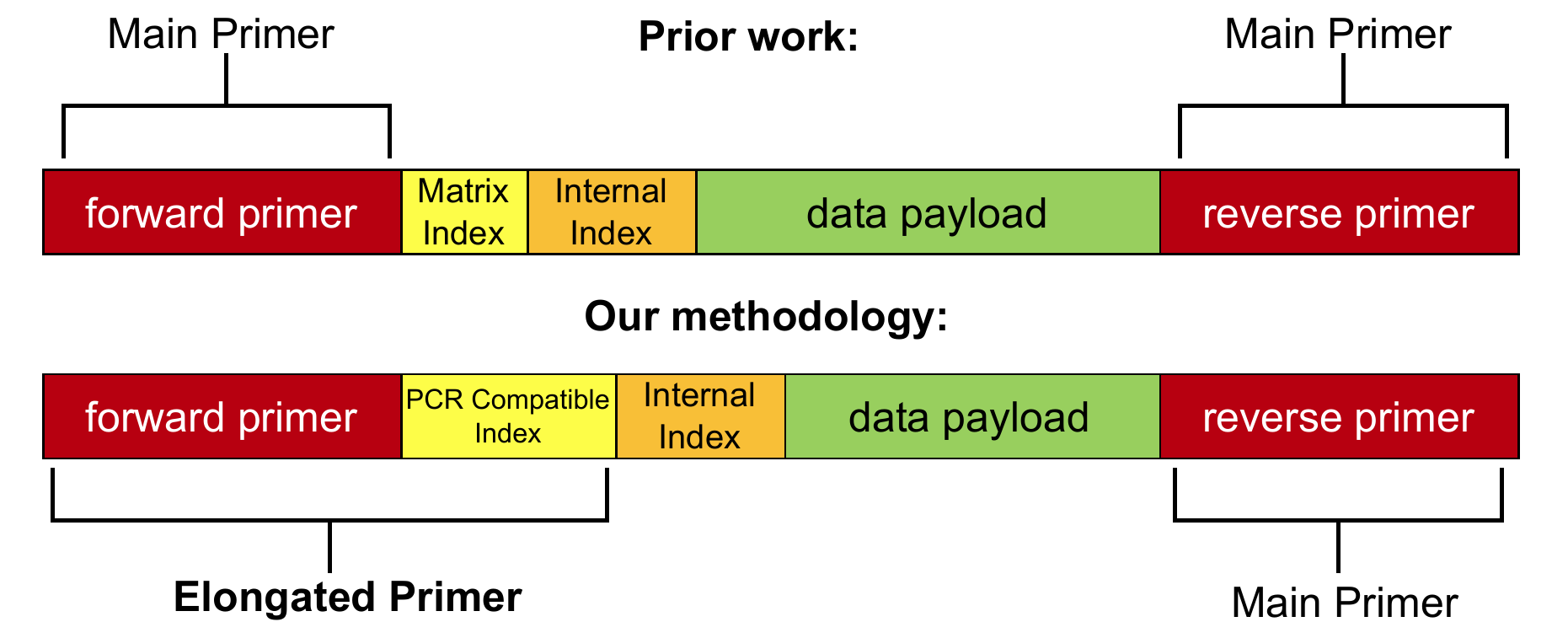}
	\caption{The structure of a DNA strand that supports PCR with elongated primers (bottom) as compared to prior work (top). Note that the primer can be elongated fully (covering the entire yellow part) or partially. PCR with partially elongated primers enable limited forms of sequential access.}
	\label{fig:elongated_primer}
\end{center}
\end{figure}

In this section we describe the methodology for creating a PCR-compatible indexing scheme for an index of length $L$. This allows us to perform precise random access with primers elongated to include a desired portion of the index. Figure~\ref{fig:elongated_primer} shows how strands with elongated primers generated following our methodology look in comparison with the baseline from prior work~\cite{organick:random}.

\subsection{General Approach}
Given that every molecule contains a unique prefix consisting of its primer and the full internal address, one could in theory retrieve each molecule individually using PCR, if the forward PCR primer included the full molecule prefix. Unfortunately, indexes AAA...AAA to TTT...TTT are not PCR-compatible, as they don't have a balanced GC content, they may have very long stretches of repeated bases (homopolymers), and, can be too similar to each other.  The main limitation of the indexes in prior work is that they follow the maximum information density design, and as a consequence we do not have any control over their structure. Our main idea is to use less dense encoding of indexes, which will give us the opportunity to introduce the desired constraints on the index structure, at a minor loss in information density. As a positive side effect, the added sparsity also provides strong protection against errors in the index, which is missing in the baseline design~\cite{organick:random}, as shown in Figure~\ref{fig:back}c. Without loss of generality, we will illustrate the procedure on an internal address space with 1024 leaf indexes, which we use for our wetlab evaluation. 
    
\subsection{Requirements for Elongated Primers}

It is important to emphasize that compared to the main primers that define a partition, our elongated primers have slightly different requirements because their usage and purpose is different. Namely, every pair of the main primers must have a high mutual distance so that we are able to extract the target partition regardless of its size and concentration relative to other partitions. Assume that we have two \emph{similar} primers, $P_a$ and $P_b$, that define two partitions A and B. If the concentration of partition A is a million times lower than B, than partition A cannot be reliably extracted~\cite{organick:random, tomek:promiscuous} using primer $P_a$ due to the sheer size of B. However, within a partition, we adhere to the constraint of uniform concentration of data within the same level of the index hierarchy. As such, although our precise PCR may amplify wrong data too, the target data is guaranteed to be dominant. As such, ensuring a high distance between indexes, although desired, is not as important as it is for the main primers.




Regarding GC content, our primers have slightly more restrictions. The reason is that our primers are not of fixed length; e.g., the main primer may be extended by 6 bases or 10 bases, and in both cases it needs to have balanced GC content to be used as a PCR primer~\cite{dieffenbach:general}. In other words, the GC content needs to be balanced within every part of every index regardless of its length, and the only way to achieve this is to add sparsity to the indexes such that the GC content is uniform within and across every possible elongation. We next describe how to add sparsity in a way that disables long homopolymer sequences, while increasing the distance between different indexes.

\begin{figure}
\begin{center}

	\includegraphics[width=0.8\columnwidth]{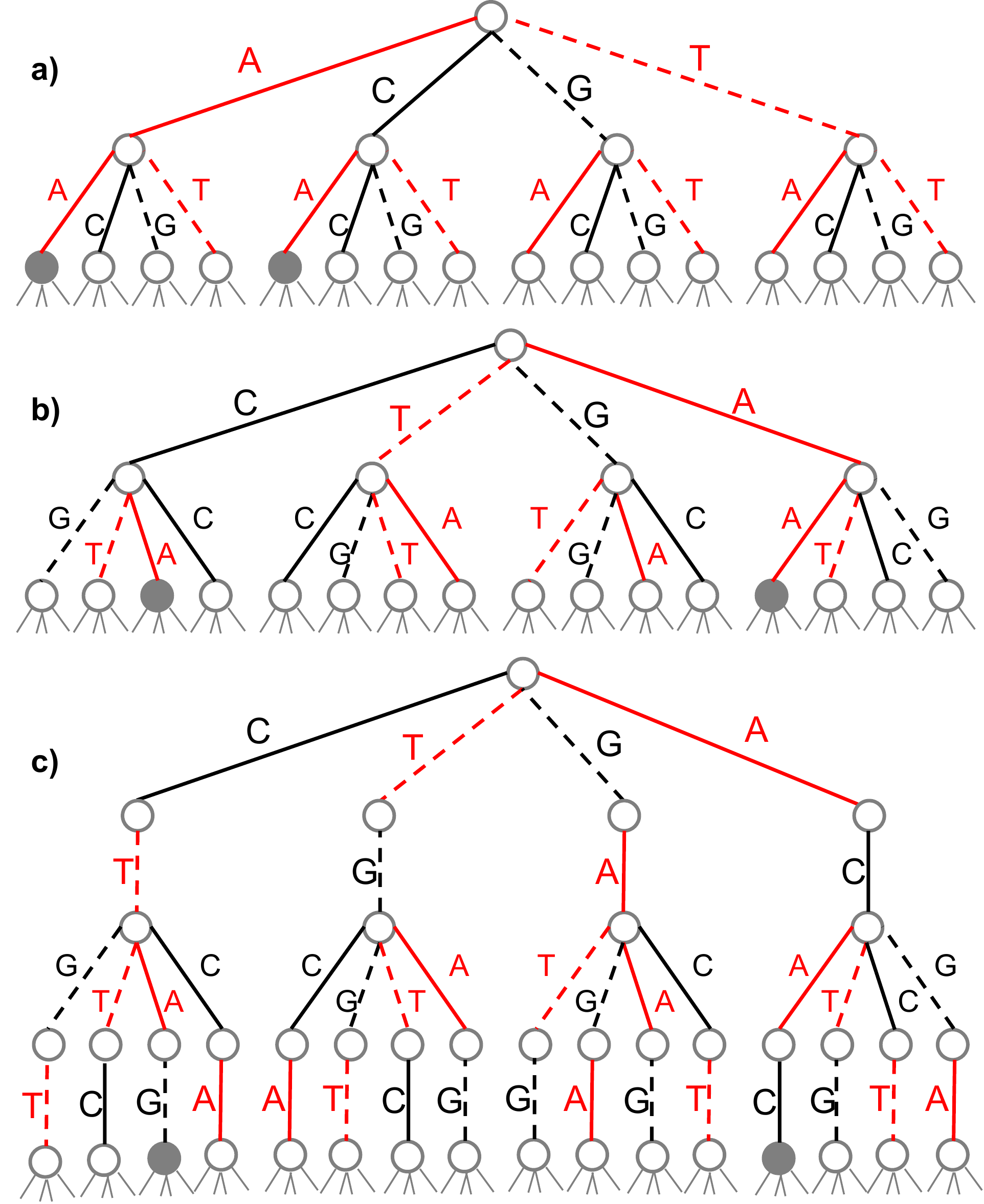}
	\caption{Index-tree construction.}
	\label{fig:tree}
\end{center}
\end{figure}

\subsection{PCR-Navigable Index Tree}
To preserve the hierarchical organization of indexes, any modification to them must be done on the prefix tree that defines their address space. Figure~\ref{fig:tree}a shows the top two levels of such a tree, where each non-leaf node has four edges labelled A, C, G, T, in that order. Path AA in this tree corresponds to address 00, in base 4; AC corresponds to 01, while TT corresponds to address 33. Before we add sparsity to this tree, we first randomize the order of edges coming out of every node, as shown in  Figure~\ref{fig:tree}b. Doing so re-enumerates indexes randomly, such that the leftmost path becomes CG and is assigned address 00. The reason for this is to support incomplete, unbalanced, and degenerate trees by avoiding situations where nodes with only one child have the single outgoing edge always be labelled as A.

Next, to sparse out the addresses, we add an extra letter between every two adjacent edges, as shown in Figure~\ref{fig:tree}c. This allows us to perfectly balance the GC content in all parts of the internal address, as we always pick a letter of the opposite GC value from the previous one. For example, if the previous letter on the path from the root was A, the extra letter could be either C or G. We pick the assignment that maximizes the Hamming distance between the sibling nodes, breaking any ties randomly. The resulting internal addressing scheme guarantees near-perfect GC content in every part of any index regardless of its length, while simultaneously disabling sequences of homopolymers longer than two. Importantly, it also increases the average Hamming distance between two indexes of the same length by at least 2x, as well as the minimum Hamming/edit distance between any siblings. In case of the two shaded nodes in Figure~\ref{fig:tree}a, AA and CA, which originally have Hamming distance of 1, after adding randomized sparsity, the two nodes become ACAC and CTAG, which have Hamming distance of 3 (Figure~\ref{fig:tree}c). 

The above indexing scheme also has cost, namely the space occupied by the added bases. In our case, we use 10-base long internal addresses of encoding units, instead of 5, which results in 3\% information density loss assuming 150-base strands. However, we believe this is a good trade-off as the added flexibility can be used to reduce sequencing costs by many orders of magnitude. Also note that this overhead reduces linearly with strand length and would present only a 0.3\% overhead with DNA strands of 1500 bases. In contrast, using 30-base primers in the baseline architecture~\cite{organick:random} would lead to 22\% loss in information density (2.2\% with 1500-base strands), and would only provide a negligible improvement in the number of chemically addressable objects.

Note that the internal address organization of molecules within an encoding unit (matrix) shown in Figure~\ref{fig:back}c (orange color) remains the same, as there is no incentive to randomly retrieve only a subset of a unit, given that the unit must be decoded as a whole. As these addresses are distinguished in software, rather than chemically, the basic addressing scheme provides the best information density for that part of the address space.

\subsection{Index Tree Management}
Because of our primary reliance on randomization and deterministic procedures in the construction of the PCR-compatible index tree, we do not need to store the tree. We only need to remember the seed used for the randomization of its construction. This seed is stored/cached digitally, along with other partition-level metadata, such as the seed used for data randomization which is used to improve clustering~\cite{organick:random, rashtchian:clustering}. We use different seeds for different partitions (i.e., primer pairs) to ensure that different partitions have vastly different trees to avoid unwanted molecular interactions between indexes of different partitions.


\section{Data Updates in DNA Storage}
\label{sec:update}

The advances in low-latency enzymatic synthesis tailored specifically for DNA storage~\cite{lee:enzymatic} and nanopore-based sequencing technologies~\cite{jain:nanopore}, as well as the introduction of near-molecule computational primitives on top of DNA storage~\cite{bee:molecular}, have expanded the potential applications of DNA storage beyond a simple read-only archival medium. Unfortunately, simple key operations such as updates are still not supported in a practical manner. In this section, we discuss some na\"ive approaches to DNA updates, and then present our proposal for updates in DNA storage with multiple possible implementations.

\subsection{Na\"{\i}ve  alternative}
A simple way to provide updates in the baseline architecture is to create a brand new, updated copy of all the data that is tagged with the same primer pair as the data that needs to be updated, then tag the new copy with a new pair of primers, and simply disregard the the old data, and notify the user/upper-level application of the new primers. While conceptually simple, this approach suffers from multiple big problems. First, it requires recreating from scratch the partition, which could have an arbitrary amount of data. Given that DNA synthesis is the most expensive process in DNA storage, this clearly is not an acceptable solution. Furthermore, the old pair of primers is wasted, as the old data with the old primers remain in the sample; given how precious primers are, wasting one pair of primers per update is clearly not a sustainable solution.

\subsection{Our Approach: Versioning}
To provide an effective solution for updates, we draw inspiration from conventional data management systems, such as journaling  file systems and systems for versioning control, such as git. Instead of in-situ chemical alteration of DNA molecules, in our proposed architecture, updates are efficiently and durably \textit{logged} as an ordered series of incremental patches. An update patch is synthesized as a minimal set of DNA molecules that describe the update to be performed on top of the original already synthesized data; these additional molecules can be easily, safely, and precisely combined with the original DNA molecules, provided that their concentration per molecule is similar. The actual application of updates is delayed until the time of decoding, at which point the updates are trivially and efficiently applied in software, obviating the need for chemically complex data edits.

The versioning approach described above is applicable to any type of data, as well as to any type of encoding, and poses no limits on the number of updates that can be performed. However, a number of important challenges remain. First, how should the updates be tagged, i.e., what should be their primers and indexes? What is the structure and semantics of update patches, and how are they applied? How to link the data and its updates? How do we efficiently retrieve the updated data? How do we physically mix original DNA with update DNA, and in what proportions? We answer these questions in the following subsections.

\subsection{Placement of updates in the address space}
\begin{figure}
	\includegraphics[width=1\columnwidth]{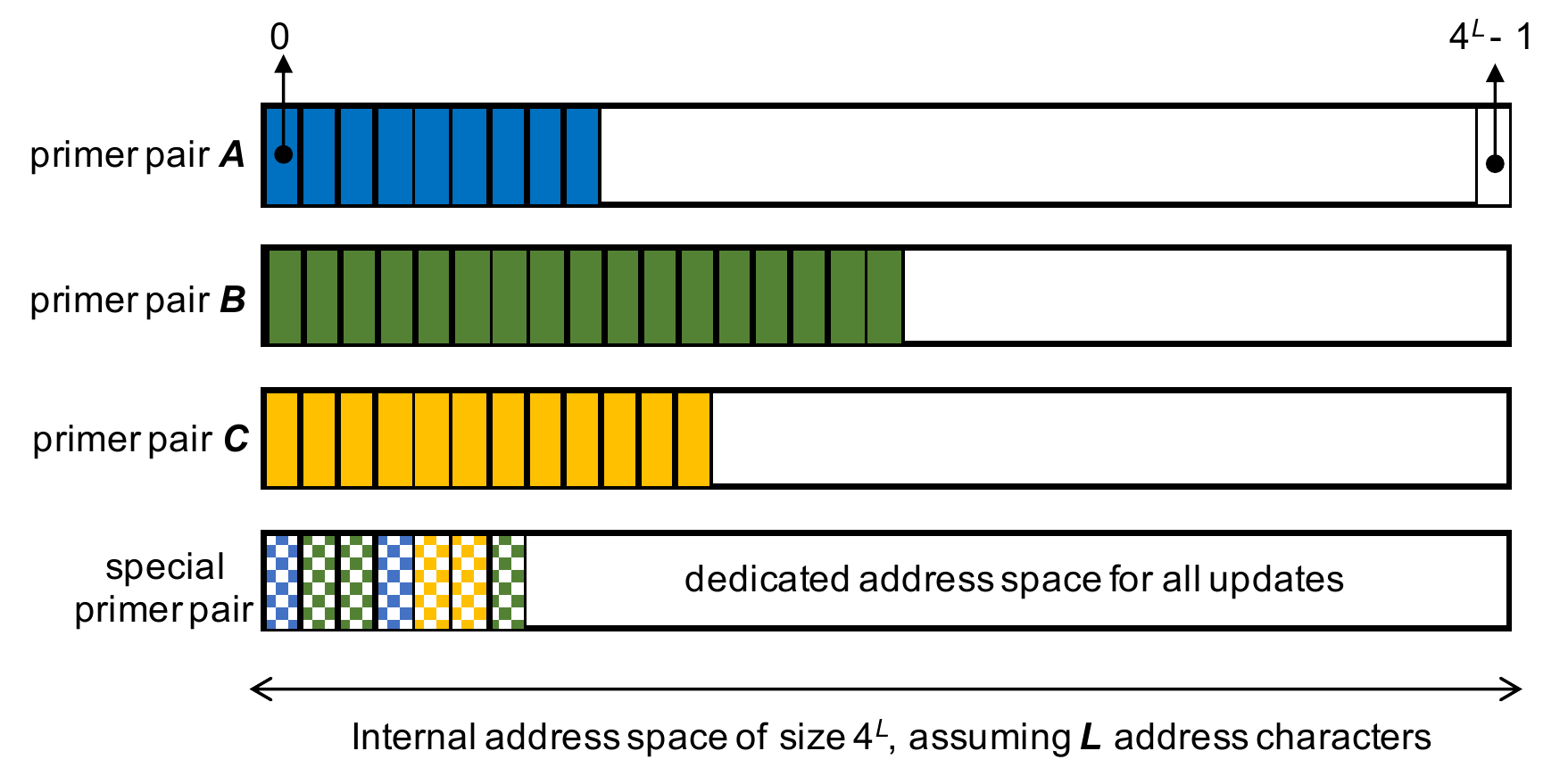}
	\caption{Logging all updates in a dedicated address space.}
	\label{fig:update1}
\end{figure}

Figure~\ref{fig:update1} shows a simple approach where all updates are logged into a separate partition defined by a dedicated pair of primers. In this scenario, all updates from all partitions are placed together in the same address space. A minor problem with this approach is that it requires a dedicated pair of primers for updates. A much more serious problem is that reading any amount of data from any partition in the same sample requires reading all the updates that have ever taken place in any partition, only because the data \emph{might} have been updated. In the worst case, reading a small piece of clean (never updated) data may require reading huge amounts of update logs, of which none are related to the target data, which is prohibitively costly and slow.

\begin{figure}
        \includegraphics[width=1\columnwidth]{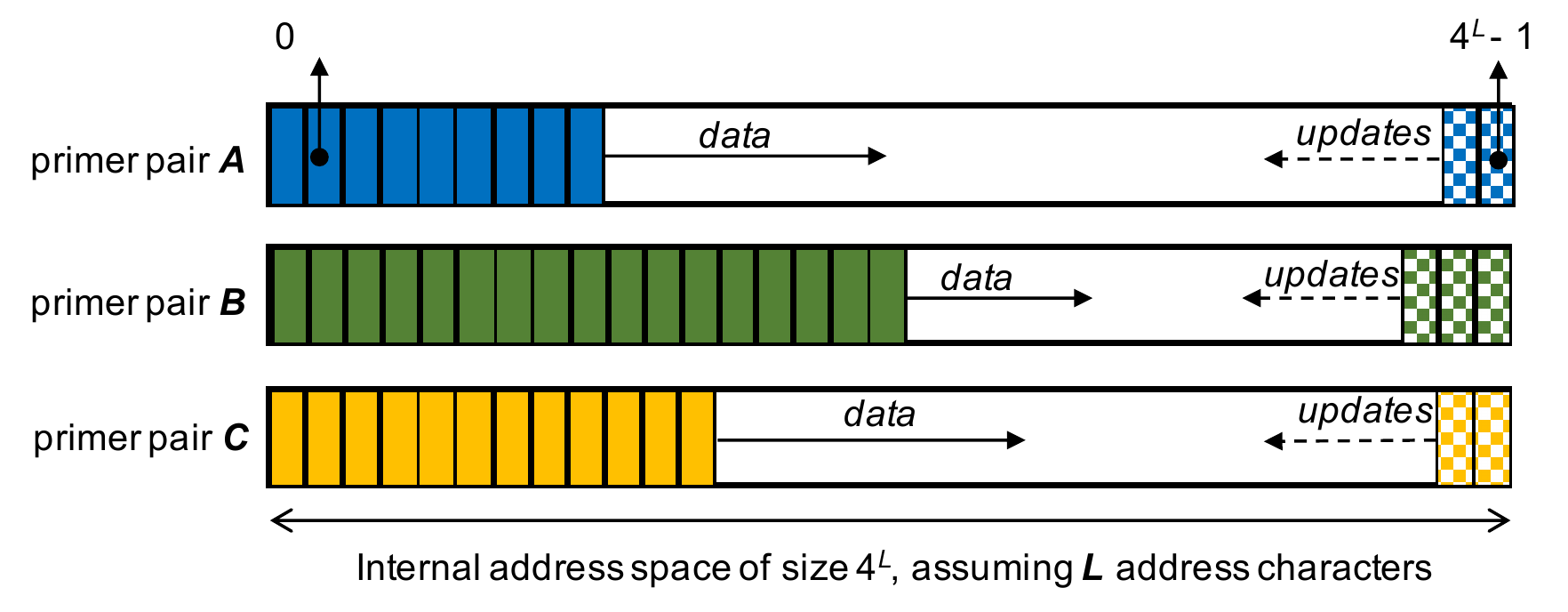}
        \caption{Logging updates in same address space as data.}
        \label{fig:update2}
\end{figure}

Figure~\ref{fig:update2} shows an improved approach, where updates are embedded into the address space of each primer pair. This way, the updates related to different address spaces are separated from each other, which means that reading updated data requires a single PCR. The data part and the update part need to share the same address space, which can be achieved in multiple ways. The most flexible organization of the address space is shown in figure~\ref{fig:update2}, where the data and updates share the address space in a fashion similar to how two stacks are placed in memory, i.e., growing towards each other. This way, there is no need to statically partition the address space into data part and update part, as both parts can grow dynamically. 

\begin{figure}
        \includegraphics[width=1\columnwidth]{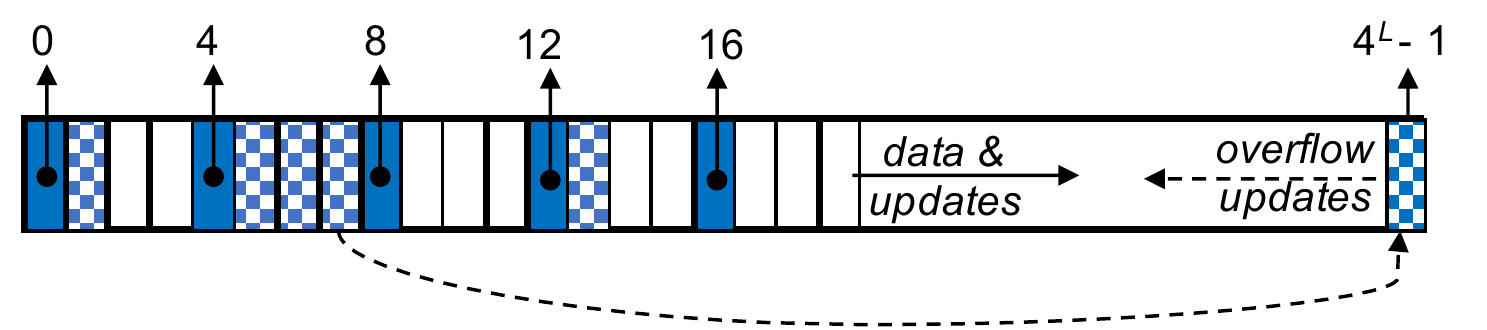}
        \caption{Interleaving data and the related in the same address space with data.}
        \label{fig:update3}
\end{figure}

While the improved approach in Figure~\ref{fig:update2} enables retrieval of both data and updates in the same PCR reaction, this retrieval would include all the data under the same primer pair, which can amount to gigabytes or terabytes in practice. Our final approach seeks to place the updates (in the address space) next to the block they are supposed to update, and far from unrelated blocks, such that a single precise PCR reaction as described in Section~\ref{sec:elongated} can retrieve the block together with all the updates that are related it. Figure~\ref{fig:update3} shows how provisioning additional slots for updates right after every data block can achieve the desired interleaving of data and updates. The method basically spaces out data blocks in the address space, to provisioning some address space for the updates. In this example, address space slots for three update blocks are provisioned for every data block. Some of these update slots could be unused, while some data may require more updates than it was statically provisioned for it in the address space. In the latter case, the last update block will contain a pointer to an entry in the common update log for updates needing more space.

The key advantage of the method in Figure~\ref{fig:update3} is that it ensures that data and the corresponding updates have a common prefix, and can be retrieved together in a single PCR. For example, assume an object with prefix ACGT. At the expense of one extra base, the system would store the original object as ACGTA, the first update as ACGTC, second update as ACGTG, etc. This establishes an implicit link between data and updates through the common prefix, and no additional bookkeeping is needed. The user always uses the same address ACGT to retrieve the object, and the system always uses that prefix to perform the PCR. Such PCR will precisely retrieve the original object and all the updates, and the system will know the correct order of updates, which is sufficient to reconstruct an updated object. Finally, by co-locating the data blocks with updates in this manner, we limit the potential size and molecular concentration imbalance between different blocks to at most 4x in this example, assuming no differential coding of updates is used. 

\subsection{Structure and Semantics of Updates}
Note that our system imposes no limitations on the structure or the semantics of updates. Given our block semantics, the updates could simply be in the form of a new block that entirely replace the old one. However, to further minmimize the size of the updates and the concentration imbalance they pose, we can encode the updates as a differential patch to the original data. In the latter case, the actual application of updates can even be delegated to the end-user or the upper-level application, which would be most efficient to do for data that is compressed in an application-specific manner. Our wetlab experiments provide one simple example of a possible update semantics that was appropriate in our miniaturized wetlab setup, but many others are possible. Because of this flexibility and because of lack of space, we do not analyze different approaches for creating and applying patches in this work. Most DNA-storage systems will have digital front-ends, which could buffer, coalesce, batch and co-schedule minor updates (as well as cache hot data/metadata).

\subsection{Physically Mixing Data and Updates}
For efficient retrieval of updated data, it is important to ensure that the original DNA data and the additional updates are mixed in adequate concentrations, such that the average number of copies of each molecule is as similar as possible in the original and update DNA strands; the mismatch in concentrations will directly impact the cost of sequencing. To understand why, let's assume that in the update sample the number of copies per molecule is 10 times higher compared to the original sample. If we directly mix such samples, then ~90\% of the sequencing output will be related to the update, and only 10\% will be related to everything else; that means that we have to sequence $\sim$10x deeper than usual to retrieve the required number of the original sequences, increasing the sequencing cost by ~10x. Similarly, if the updates are five times less concentrated than the original, sequencing the updates from the mix would again require about 5x higher sequencing coverage.

Ensuring the adequate concentrations may seem challenging, as the original DNA and the update patch may come from different DNA synthesis vendors, using vastly different synthesis technologies, with different yields and concentrations. For example, the DNA patches that we ordered were small in terms of the number of distinct molecules and as such they were much cheaper to synthesize using a different vendor. The new vendor, however, provided samples that were 50000x more concentrated compared to the samples with the original data. Another issue is that the original sample (or the update) may have undergone a series of PCR amplification from the moment of synthesis, making the problem of concentration matching even more challenging. Fortunately, the most basic wetlab tools and simple calculations allow us to match the concentrations with remarkable precision. In subsection~\ref{sec:54}, we describe two simple and successful protocols for mixing the samples.

\section{Methodology}
\label{sec:methodology}
As a proof of concept, we demonstrate the effectiveness and precision of our proposed read method with elongated primers, as well as the update mechanism, through wetlab experiments. The experimental details below are provided for reproducibility, and we also plan to release our sequencing data to support research in this area.

\subsection{Input Data and Experiments}
We encode 13 files into 13 partitions; 12 of these files simply present unrelated data partitions in the same DNA pool. The last file is the book \textit{Alice's Adventures in Wonderland} by Lewis Carroll, 150KB in size, encoded with a separate pair of primers. The index space of the last file is organized to be PCR-compatible with 1024 leaf nodes, as described in Section~\ref{sec:elongated}. This file is split into about 600 equal encoding units (blocks), each having 15 molecules, 4 of which are used for ECC. The binary size of each encoding unit is 256 bytes, which is about the size of an average paragraph of text in the book, and each unit is assigned to one leaf sequentially.

All 13 files were synthesized into DNA by Twist BioScience. We then retrieve individual files as well as precisely retrieve certain paragraphs from \textit{Alice's Adventures in Wonderland} using elongated primers. We also achieve successful retrieval of multiple unrelated paragraphs together in one multiplex PCR reaction. We then update some paragraphs and retrieve them after the updates. Although our experimental setup is miniaturized to allow for an inexpensive demonstration, these experiments can be easily reproduced in significantly scaled-up setups, where each paragraph can represent an arbitrary type and amount of information. In our experiments, each paragraph represents one block.

\subsection{Storage Architecture}
Our baseline DNA architecture is modelled after the state-of-the-art architecture that has successfully demonstrated random access among 200MB of data in DNA~\cite{organick:random}. We use 150-base long DNA strands~\cite{organick:random}, as we could synthesize those in the most cost-efficient manner given the scale of our experiments. Out of 150 bases, 40 were used for a pair of main access primers. One \textit{A} base was added after the forward primer as a point of synchronization~\cite{organick:random}, leaving 109 bases for data (or ECC) and internal addresses.

The encoding unit size is defined by the size of the Reed-Solomon symbols used. To reduce the cost of experiments, we use small 4-bit symbols, which means that a codeword has $2^{4}-1 = 15$ symbols, i.e., the matrix in Figure~\ref{fig:back}c has 15 columns, 11 of which are data  molecules as in Figure~\ref{fig:back}a, and the remaining four are ECC molecules as in Figure~\ref{fig:back}b. The data part of a molecule contains 96 bases, which is 24 bytes, so the entire encoding unit contains 264 bytes, 256 are used for data and the remaining 8 bytes are randomly padded. For more details on the ECC, please refer to the prior work~\cite{lin:managing,organick:random}.

\subsection{Indexing}
For addressing \emph{within} an encoding unit (i.e., matrix), we need only two bases for that part of the index (the part colored in orange in Figure~\ref{fig:back}), from AA to GG, which is enough to distinguish between 15 molecules in software. Although with our 600 encoding units we need only 5 bases to densely encode the address space of 1024 units, we use 10 bases to create a sparse GC-balanced and PCR-friendly internal address of the encoding unit (the part colored in yellow in Figure~\ref{fig:back}) and one base to support updates. 

\begin{figure*}[htp]
\centering

\subfloat[random access for the whole partition]{%
\includegraphics[clip,width=0.6\columnwidth]{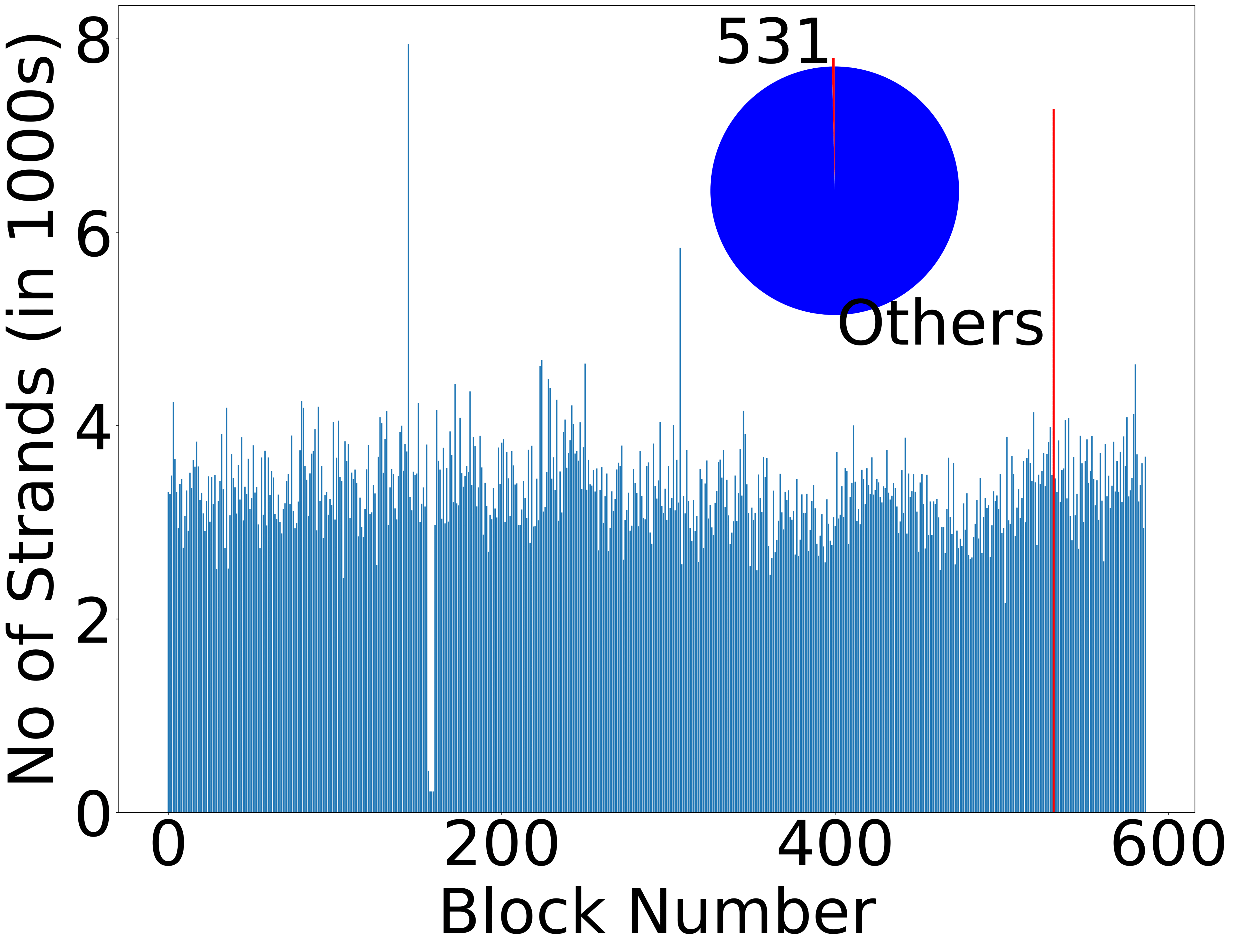}%
  \label{fig:twist}
  }\qquad
\subfloat[random access for block 531]{%
  \includegraphics[clip,width=0.6\columnwidth]{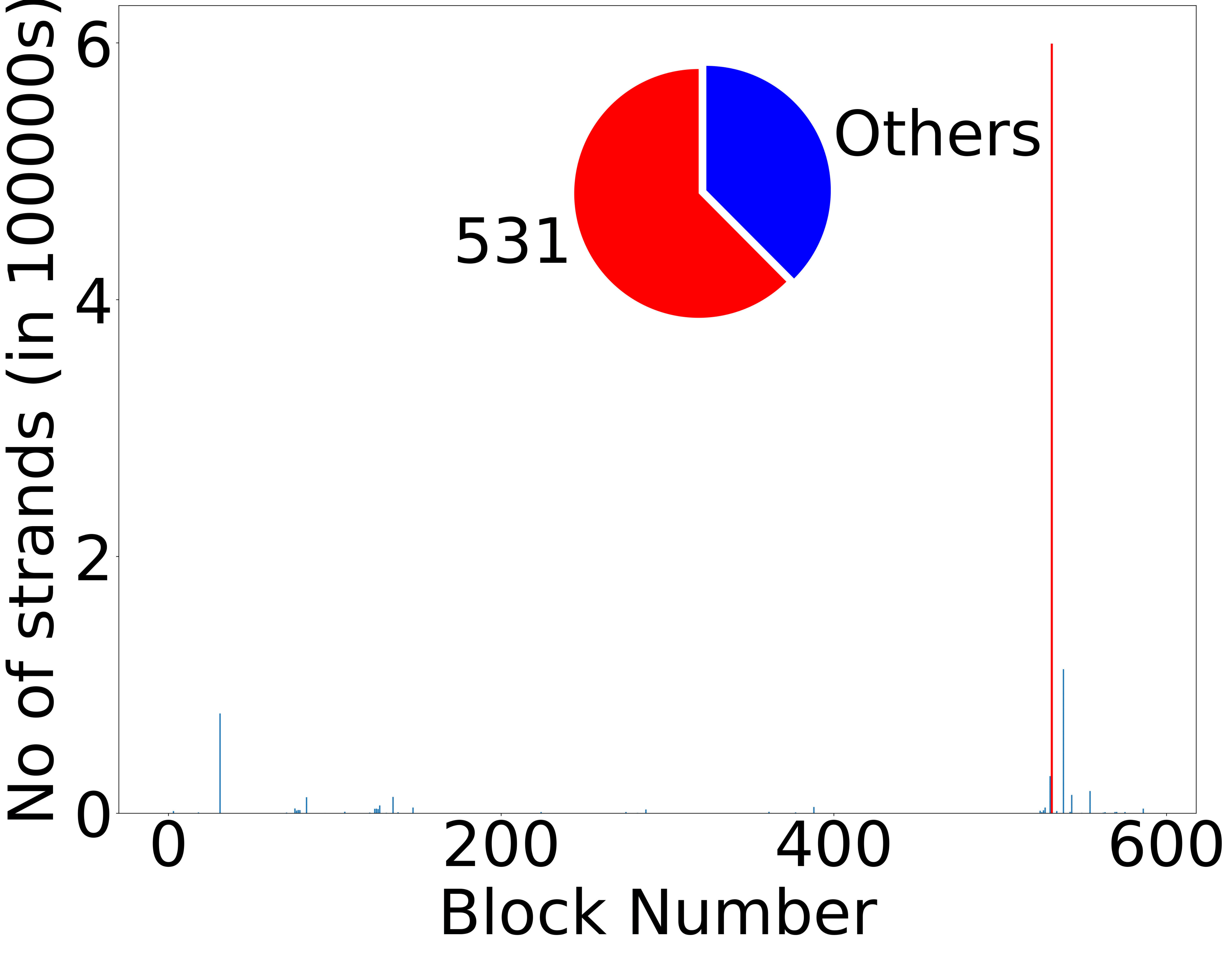}%
  \label{fig:clean}
}\qquad
\subfloat[random access for block 144]{%
  \includegraphics[clip,width=0.6\columnwidth]{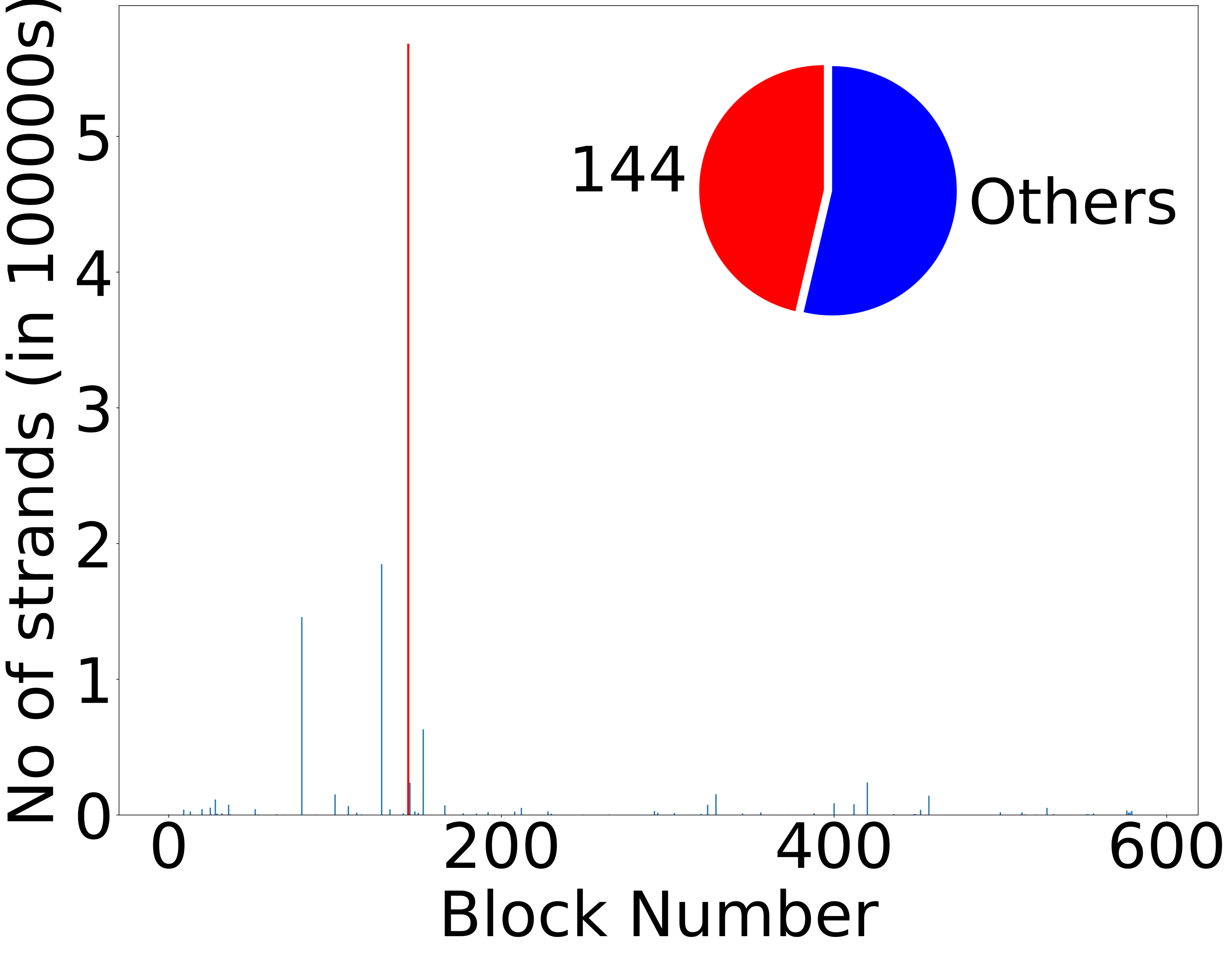}%
  \label{fig:clean144}
}

\caption{Distribution of blocks in the sequencing output after PCR-based random access with different front primers a) using the main partition primer, b) using the elongated primer for block 531, c) using the elongated primer for block 144.}
\end{figure*}

\subsection{Updates}
Every update is encoded as an encoding unit (matrix), similarly to the data. In our proof-of-concept setup, our format of the updates is very simple and consists of four parts. The first byte is an integer that identifies the first byte in the block (encoding unit) where deletion needs to happen. The second byte is a number that indicates how many bytes, starting from the one indicated by the first number, should be deleted, if any. The third part contains an integer that identifies the position of where an insertion should happen, after the deletion is applied. The rest is an array of bytes that should be inserted. Note that the block that needs to be changed is identified by the internal address of the update patch: the data and the update only differ in the last base.

\subsubsection{Wetlab experimental setup}

We synthesized six update patches in total, to update six blocks. Three of them were synthesized by Twist BioScience together with the original DNA, in the same pool. The other three updates were synthesized by a different company, Integrated DNA Technologies (IDT), as a separate DNA pool containing 45 molecules, because the synthesis of a small update pool was significantly cheaper at the second company. However, the IDT pool was 50000x more concentrated compared to the original Twist pool; the two pools therefore required careful mixing, which we successfully achieved with remarkable precision.

\subsubsection{Physical mixing of data and update pools}
\label{sec:54}
We performed two different mixing protocols targeting two use cases. In one of them we try to match the concentrations while taking data directly from the synthesized pools. In the other, we first amplify the pools and try to match the concentrations after amplification, to simulate the situation where the original synthesized pools are not available.  

In the first approach, which we name \emph{Measure-then-Amplify}, we generate the mix by measuring the concentration of the unamplified Twist pool and IDT update pool and mixing the two pools together with appropriate dilutions to ensure that an equivalent amount of IDT sequences is added to the Twist pool. These were then subsequently amplified (95$^{\circ}$C denaturation, 55$^{\circ}$C annealing and 68$^{\circ}$C extension, 15 cycles), with the main partition primers to generate the mixed pool.

In the second approach, which we name \emph{Amplify-then-Measure}, we initially amplify via PCR the initial Twist oligo pool using the main partition primers. PCR amplification was conducted for 15 cycles (98$^{\circ}$C denaturation, 55$^{\circ}$C annealing, 68$^{\circ}$C extension). PCR cleanup was performed (using the Geneaid Small DNA Fragments Extraction Kit). IDT oligo pool was similarly amplified and cleaned up. Concentrations of both pools were measured via nanodrop, before mixing them both in concentrations proportionate to the number of unique oligos in each pool (8850 for amplified Alice pool and 45 for IDT update pool), ensuring an even distribution of oligonucleotides without  under- or over-representation of the updated pool.

The evaluation of the mixing protocol is shared in subsection~\ref{subsubsec:mixing}. It should be noted that in this work we have merely used basic chemical methods to clean samples and measure concentration. There are many advanced ways to improve upon the efficacy of our processes by using better PCR cleaning techniques~\cite{lin:dynamic} and more precise concentration measurements~\cite{organick:probing}.

\subsection{Performing Random Block Access}
Elongated forward primers (31-base long) for blocks 144, 307 and 531 were used together with the regular reverse primers (20-base long) for initial amplification from the pre-amplified Alice partition. Four PCR reactions were conducted. The first three utilized 144, 307 or 531 primers, one at a time, while the last utilized an equal mix of all three for multiplexed amplification, with the total primer concentration of the mixed pool being the same as in the case of the single primer pair. Touchdown PCR was conducted to increase the specificity of the amplification process, with a decrease of 1$^{\circ}$C per annealing step in each cycle, starting at 65$^{\circ}$C, for 10 cycles, before amplification at 55$^{\circ}$C and annealing for another 18 cycles. The GC content of all primers is between 48-52\%. The melting temperature of the elongated primers is between 63-64$^{\circ}$C.
To prepare the data for sequencing, Illumina sequencing adapters were synthesized as primers and appended to all sequenced pools via overlap-extension PCR. Unique indexes were also added in this process to each sample for demultiplexing. The final product was verified through Sanger sequencing and subsequently sent for commercial sequencing with Illumina NovaSeq.

\subsection{Decoding}
The sequencing output of every experiment is first clustered using a specialized clustering algorithm developed for DNA data storage~\cite{rashtchian:clustering}. The resulting clusters were run through a two-sided consensus finding algorithm~\cite{organick:random}, after which every molecule is placed into an appropriate matrix, based on the internal address of the matrix (yellow part in Figure~\ref{fig:back}) and the internal address within the matrix (orange part in Figure~\ref{fig:back}). Every matrix is then decoded and error correction is applied. Finally, updates, where they exist, are trivially applied as per the instructions found in the decoded update patches. A more detailed decoding procedure in the context of our results is described in section \ref{sec:decod}.

\section{Results}
\label{sec:evaluation}









\subsection{Baseline Random Access}
Figure~\ref{fig:twist} shows the number of reads in each block in the original pool after a simple PCR random access using the main partition primers, which among the 13 files amplifies only file 13 (Alice in Wonderland). This is the conventional random access as per the baseline~\cite{bornholt:dna, organick:random}. We show this figure to emphasize the fact that prior to any precise reads, all molecules are represented fairly uniformly in the DNA pool, with minimal bias (within 2x) coming from either synthesis or PCR processes. Also note that three of the blocks (144, 307, and 531), seem prominent because they have about twice as many molecules compared to other blocks, as they contain both the original data and the updates that were synthesized together.

\textbf{Cost Implications.} Assuming that the user wants to retrieve updated block 531, the above approach produces a sequencing output with 99.66\% of unwanted data, and only 0.34\% of wanted data, i.e., block 531. As the sequencing cost is charged per byte of sequencing output, 99.66\% of that cost will be wasted due to poor selectivity of the random access. In other words, to retrieve $x$ amount of block 531, the baseline system has to sequence $1/0.34\% = 293x$ of unwanted data. This cost can only be reduced if the sample sent for sequencing includes less of unwanted data, i.e., if the target data is more selectively amplified.



\subsection{Random Block Access}
Figure~\ref{fig:clean} shows the results after precise random access for block 531 (plots for other amplified blocks look similar and are omitted for brevity). Before plotting this figure, around 18\% of reads were discarded as they were amplified by the leftover main primers from the previous reaction. The remaining 82\% had the correct target prefix for 531. However, only 59\% of those actually are copies of block 531; the remaining 41\% originate from a handful of other blocks that were promiscuously amplified through mispriming, as their index partially overlaps with the index of 531. Section~\ref{appendix:prom} discusses how to detect and handle incorrect amplification. Nevertheless, the target amplification was strong enough that the target can be clearly identified, with $0.82 \times 0.59 = 48\%$ of the reads mapping to the target.

\subsection{Sequencing Cost Reduction}
The ability to retrieve individual data blocks has a huge impact on the sequencing cost. Conceptually, the sequencing cost is  always proportional to the size of the sequencing output, regardless of the sequencing technology. Thus, sequencing a specific block reduces the cost linearly as compared to sequencing the entire partition. 

In our specific experiment, to read $x$ amount of block 531 (at any desired coverage), the baseline system has to sequence $1/0.34\% = 293x$ of unwanted data, while in our case only $1/0.48 -1 = 1.08x$ of the sequenced data is unwanted. This reduces the sequencing cost $(293+1)/(1.08+1) = 141$ times, regardless of the sequencing technology used! Our experimental results match these calculations and are discussed further in section \ref{sec:decod}.

\subsection{Sequencing Latency Reduction}
Apart from reducing the cost of sequencing, the ability to select a given block also reduces the sequencing latency. The reduction in latency, however, can be dependent on the partition size and the sequencing technology used, which we further explain.

The duration of a single next-generation sequencing (NGS) run is fixed by design. For these machines, the sequencing output is available only at the end of the sequencing run, which takes a fixed amount of time and produces an output containing a fixed number of reads. Thus, for small partition sizes that fit into a single sequencing run, the reduction in the sequencing latency is conceptually impossible. However, if the amount of data in a given partition is huge, multiple sequencing runs may be required to sequence all the molecules at a sufficient coverage in order to decode the entire partition. For example, one run of Illumina MiSeq can only produce around 1GB of user data in the best case. Sequencing a partition of 1TB would therefore require $\sim$1000 runs. However, in our block-based architecture, retrieving a single block instead of the whole partition would proportionately reduce the number of sequencing runs needed (in our case $\sim$141 times), providing a linear reduction not only in the cost, but in the retrieval latency as well.

In case of Nanopore sequencing, runtime of a \emph{single} sequencing run is always output-size-dependent and ranges from several seconds to several days; the output is continuously produced and analyzed in real-time, and the sequencing can be stopped once the data is successfully decoded~\cite{wang:nanopore}. Therefore, the latency of retrieving a block through Nanopore sequencing would always be reduced linearly compared to retrieving the entire partition, regardless of the partition size. In our case, this means that the latency of retrieving block 531 would be reduced $\sim$141 times compared to the baseline that retrieves the entire partition.

Regardless of the sequencing technology used, the post-sequencing data movement and software decoding time will also be reduced to a certain extent in the case of block-based DNA storage. However, we do not quantify those as they do not represent a bottleneck in the retrieval process~\cite{ceze:molecular}.

\subsection{Cost of Creating and Retrieving Updates}
For the baseline system that supports updates, we assume the na\"{\i}ve system described in Section~\ref{sec:update}, which creates a new updated copy of the entire partition and assigns a new primer to it. We evaluate two aspects of the updates: the synthesis cost of performing the update on a block, and the sequencing cost of reading an updated block. 

The cost of updating block 531 in the baseline includes synthesizing the entire new partition (8805 molecules), whereas in our system it requires the synthesis of 15 molecules of updates, which is a reduction of approximately 580$x$. To read the updated block 531 at any given coverage, the baseline system must read the entire partition and discard most of the readout, whereas our system can perform the precise access that retrieves both data and updates related to block 531 (30 molecules in total) as shown in Figure~\ref{fig:clean}, discarding only about 50\% of reads and reducing the sequencing cost for updated data by approximately $0.5*(8805/30) = 146$x.

\subsubsection{Other Costs}
Note that, apart from the synthesis and sequencing costs, the na\"{\i}ve system has other hidden costs. First, it reduces the storage density of the system by keeping a full copy of both the old and the new data. Second, it requires a new pair of main primers for each update.  Finally, it must notify the end user or the upper layer in the hierarchy about the change of primers of the updated object. Our solution successfully eliminates all of these costs.

\subsection{Mixing Data and Updates}
In our experiments, we attempted to update 6 of the 587 blocks in the DNA pool. We were able to successfully retrieve the data with all 6 blocks updated, including the 3 updated blocks in the IDT synthesis batch that were synthesized later and mixed with the original data. 
\begin{figure}
	\includegraphics[width=1\columnwidth]{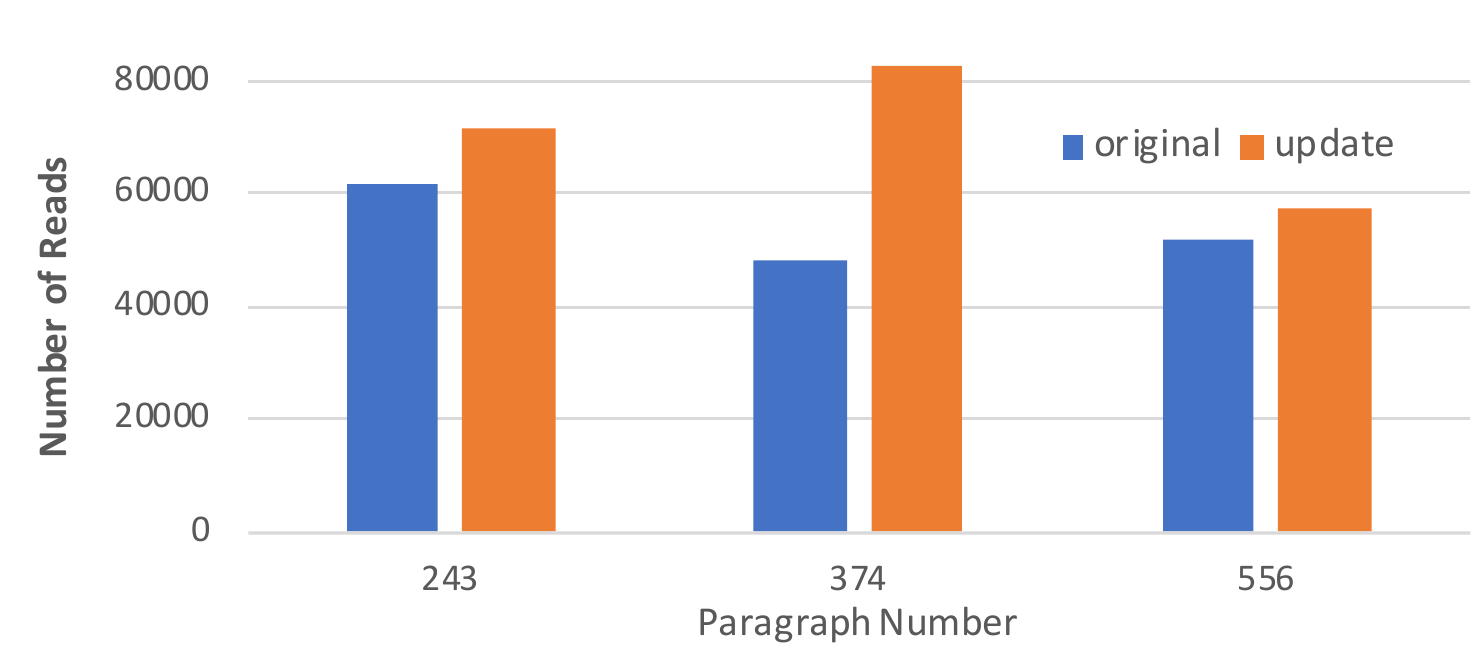}
	\caption{Mixing Outcome: The number of original and update molecules in paragraphs 243, 374, and 556 after mixing with the original pool using Amplify-then-Measure.}
	\label{fig:amc}
\end{figure}

\label{subsubsec:mixing}
 Figure~\ref{fig:amc} compares the number of original and update molecules in paragraphs 243, 374, and 556, which were updated separately, after mixing them with the original pool using the Amplify-then-Measure approach (the Measure-then-Amplify numbers are similar and thus omitted for brevity). In both approaches, we were able to mix the original and the update pools in a way that matches the concentrations quite well, despite their enormous initial difference in concentrations, with some natural variation as seen in Figure~\ref{fig:twist}. This demonstrates the ease and practicality of mixing two samples in appropriate concentrations with high precision, which allows us to keep the cost of sequencing for updated data at the minimum.  

\subsection{Scalability and Limitations}
\subsubsection{Block Count}
Our wetlab results show that elongation of the front PCR primer can provide at least $1024$ blocks that can be individually addressed with a tolerable level of mispriming. However, when using our proposal, one should opt for extending both primers instead of one. Such an approach would require splitting the index into two parts, and each primer would be extended by half the length of the index. We believe that the PCR efficiency and specificity would be significantly improved due to lower and more balanced melting temperatures of the two primers~\cite{basu:pcr}. Note that two-sided extension by 10 characters would create over a million $(1024^2)$ addressable blocks in a single partition, which is of the same order of magnitude as the number of pages in memory or blocks in modern SSDs. The reason we demonstrated single-sided extension is that we wanted to stress-test and analyze long primer extensions in a cost-efficient manner. 

\subsubsection{Block Size}
An important observation is that the amount of mispriming does not depend on the block size; mispriming is dependant only on the number of blocks and the structure and sparsity of their indexes. While the block size in our evaluation is only 256B, it is important to note that there are virtually no limits to its size. In fact, the way to scale the partition capacity is by scaling the block size. 

\subsubsection{Partition Count}
Directly performing a PCR reaction with an elongated PCR primer using a sample that contains many partitions may cause additional mispriming, as some indexes from unrelated partitions could be accidentally similar and may bind together. To avoid this problem, we suggest a two-stage PCR protocol, in which we first add the main primers to isolate the target partition, and after a number of cycles, add the elongated primer and continue with the PCR to isolate the target block. A similar approach has been done with nested primers~\cite{tomek:driving}.

\subsubsection{Management of Elongated Primers}
In our experiments the elongated primers were synthesized from scratch by a commercial vendor because of convenience and price. However, in a production system the synthesis of elongated primers can be done by continuation of synthesis on top of the existing main primer, saving on time and cost. Note that we suggest lazy, on-demand primer elongation, i.e., only for blocks that are actually demanded, rather than upfront for all possible blocks. In all storage systems the popularity of objects follows the Zipfian distribution, where many blocks are never accessed, and a few are accessed very frequently. Those accessed frequently will pay the price of primer elongation only once and amortize the cost over the subsequent requests. A potential problem is the physical management of too many primer elongations. One solution is to limit the number of elongated primers stored per partition (e.g., keep up to N most frequently requested elongations per partition, discard the rest).

\section{Decoding Procedure}
\label{sec:decod}
After performing PCR using elongated primers, we can recover our target block with very few reads. With just 225 sequenced reads, we successfully decoded both the original block and the updated block using the following procedure: 1. We first search for the elongated forward primer and reverse primer of our target block in our reads and extract the substring between them as the payloads.
    2. We then cluster these payloads as per Rashtchian et al.~\cite{rashtchian:clustering} so that the payloads from the reads of the same original strand are clustered together.
    3. In the descending order of the cluster sizes, we perform trace reconstruction using double sided BMA algorithm as described in Lin et al.~\cite{lin:managing} until we have a reconstructed strand for every address within the original block and any updated version, if present. We discard any reconstructed strand that has the same address as a previously recovered strand.
    4.  Once we have a reconstructed strand for every address within a block, we can decode our original data from these reconstructed strand. If there are any errors remaining, we can correct them by utilizing the error correcting codes.

In our experiment, we targeted and recovered block 531 in file 13. This block had an original version and one update, which were both recovered. Both the original and updated version had 15 strands each. In order to recover all 30 strands, we had to perform trace reconstruction on the first 31 largest clusters. The reconstructed strands were 100\% accurate, and no error correction needed to be used to fully decode the data. In contrast, to recover the same block and update prior to performing precise PCR, we needed to sequence around 50000 reads, out of which only $30/8850 = 0.34\%$ are useful reads of our target block and its updates. (We have 8805 strands for 587 blocks along with 45 strands for three updates in our solution, prior to PCR. Our target block and update consists of just 30 out of these 8850 strands.)


\subsection{Handling of Mispriming during PCR}
\label{appendix:prom}
While during our experiment we were able to recover the target blocks using our decoding procedure, it should be noted that it can be thrown off by the process of mispriming. When we perform PCR using our elongated primers, not all the strands amplified are from our target blocks. As shown in Figure~\ref{fig:clean}, a fraction of the amplified strands belong to other blocks that have had their primers overwritten by the target primer, but they retain their original payloads.
Upon examining these promiscuously amplified strands, we observed some common properties that we should avoid when designing indexes.

The incorrectly amplified strands largely had indexes that were very close to the indexes of our target block in edit distance, rather than Hamming. They were observed to be usually 2 or 3 edit distance apart. The ease of decoding a block mostly relates to the number of other indexes within this edit distance radius.

We observed that when the target block had an index that was close to the indexes of very few other blocks in edit distance, a smaller percentage of the amplified strands were incorrect. Such blocks are relatively very easy to recover and require very few strands to be sequenced for full recovery.




In a situation where the target block has a strand with very few strands in the initial solution, there is a possibility that after PCR, clustering and reconstruction, a misprimed strand might be mistaken for the original strand. This risk increases if the misprimed strand had higher concentration in the original pool. In this scenario, the target block would have a recovered strand that is entirely incorrect.

To an extent, such an error can be corrected using error correction codes. However, if the sequencing is of poor accuracy and we wish to preserve our ECCs to handle other types of errors, we can use a more computationally expensive method to handle this situation. In step 3 of our decoding procedure, we can keep reconstructing the strands from the largest clusters in descending order of cluster size, until we have multiple candidates for some of the indexes. Once we have these candidates, we can recursively try to decode the original data using each of these candidates, until we correctly recover our data. This is computationally inexpensive if the number of candidates and number of incorrectly recovered strands remains small.


\section{Related Work}
\label{sec:disc}

 
Increasing the number of unrelated addressable units (partitions) is a very important problem in DNA storage. A recently proposed solution uses \emph{nested} PCR, in which two forward primers are encoded in every strand and two PCRs are performed back-to-back, with the 2nd forward primer inserted in-between~\cite{tomek:driving}. This is achieved with a minor accuracy loss and a loss in information density. Another recent work allowed objects to share one primer (with some restrictions), but not the other, which increases the number of unrelated addressable units at a minor loss in PCR accuracy~\cite{winston:combinatorial}. Both of these approaches are orthogonal to and compatible with ours, as they seek to increase the number of partitions, while we look at the partition architecture. Unlike our approach, none of these works support multiplex-PCR, because the same primer is shared by many files. One advantage of both prior approaches is that a reasonably sized library of primers can be pre-synthesized, and when reading data, combinations of these primers can be used for PCR to access different files. In our approach, each block has its own unique elongated primer, which is synthesized on demand. The large number of individual blocks in our approach means that it is unrealistic to maintain a pre-synthesized library for all of them.

Apart from increasing the number of addressable objects, nested primers~\cite{tomek:driving} effectively create a two-level hierarchy. While this approach results in less noise compared to ours, its synthesis overhead in terms of the number of extra bases is 4x higher than ours (we need 5 extra bases for the sparse index, whereas nested PCR requires 20 bases for an extra primer). At the same time, our approach establishes a deep and narrow six-level hierarchy between the blocks that gives us the ability to perform sequential access, with only 5 added bases. The 5 added bases further contribute to the reliability of the index. To achieve the same depth of the hierarchy, nested primers would need 6 front primers, which would reduce the information density by at least 10x in our setup with strands of length 150. However, the advantage of nested primers is that each addressable unit can be arbitrary in size.

Elongation produces an exponential number of block addresses. For example, our elongation by 10 bases produces $2^{10}$=1024, similar to one level of primer nesting that requires 20 bases. Thus, in terms of sheer addresses per base, elongation produces more of them. However, these addresses are not directly comparable, as elongation-produced addresses must map to a fixed amount of data (convenient for blocks), whereas nesting can handle arbitrary data sizes (convenient for partitions). Thus, if a very high number of partitions is needed, we suggest nesting, whereas elongation should be used for dividing a partition into blocks.


Recent work leverages molecular promiscuity by encoding one set of data with the main 20-base primer, and an additional set of data with a different primer, minimally-distant from the main primer~\cite{tomek:promiscuous}. By tweaking the PCR conditions the authors show that it's possible to amplify data tagged with the minimally-distant primer on top of the main data, implementing a subset/superset relationship. A problem with this approach is that the data targeted by promiscuous PCR conditions must be present in significantly larger concentration compared to data targeted by the stringent PCR conditions~\cite{tomek:promiscuous}, which significantly reduces the information density. Additionally, it may be impossible to distinguish between the two sets of data given the similarity of their primers, as the main primer may overwrite the similar primers during PCR, as it was the case in our experiments. However, our work could benefit from the control of PCR  stringency by adjusting the concentration of magnesium chloride and potassium chloride~\cite{tomek:promiscuous} to reduce the noise.



\section{Conclusions}
\label{sec:conclusion}

In this paper, we proposed efficient support for block storage semantics with updates in DNA-based data storage, as well as some forms of sequential access. We showed that the internal address space given to any pair of PCR primers can be organized in a way that enables random access to smaller units of fixed size. This is achieved by extending the forward PCR primer to include a part of the internal address to narrow the scope of the reaction, avoiding replication of undesired data and reducing the cost of sequencing by orders of magnitude. We also demonstrated that updates in DNA storage can be easily supported using an approach similar to versioning in conventional systems. In our proposed system, the updates are durably logged as synthesized DNA patches, which can be safely and precisely mixed with the original data, delegating the application of updates to software and avoiding  chemically complex and risky direct edits of DNA molecules. Finally, we demonstrated that updates can be carefully integrated into the internal address space in a way that allows for retrieval of desired data and the related updates in one PCR reaction, which is particularly important for slow storage media such as DNA, where an extra level of indirection cannot be afforded.

\begin{acks}
The authors would like to thank Karin Strauss from Microsoft Research for her valuable comments during early stages of this work, as well as the anonymous reviewers for their constructive feedback. This research was supported by the Advanced Research and Technology Innovation Centre (ARTIC) at the National University of Singapore under grant FCT-RP1 A-0008129-00-00, and by the Ministry of Education in Singapore grants A-0008143-00-00 and A-0008024-00-00.
\end{acks}

\bibliographystyle{ACM-Reference-Format}

\bibliography{paper}


\end{document}